\documentstyle[prl,aps,multicol,epsfig]{revtex}
\renewcommand{\narrowtext}{\begin{multicols}{2}
\global\columnwidth20.5pc} 
\renewcommand{\widetext}{\end{multicols}
\global\columnwidth42.5pc} \multicolsep = 8pt plus 4pt minus 3pt

\begin{document}
\draft
\title{Photon assisted tunneling in a superconducting SET transistor 
induced by the Josephson oscillations of a SQUID} 
\author{E. H. Visscher, D. M. Schraven, P. Hadley,
and J. E. Mooij}
\address{Applied Physics and DIMES, Delft University of Technology\\
Lorentzweg 1, 2628 CJ Delft, The Netherlands}
\date{\today}
\maketitle
\begin{abstract}
Josephson oscillations generated by a SQUID were used to measure 
photon assisted tunneling in a superconducting single electron tunneling 
(SET) transistor. The SQUID was fabricated only a few microns from the 
SET transistor. The close proximity of the generator to the SET transistor 
allowed for a continuous sweep of the excitation frequency from 10 GHz 
to 190 GHz. The amplitude of the Josephson oscillations were tuned by 
adjusting the flux through the SQUID loop while the frequency was 
independently adjusted by controlling the bias voltage across the SQUID. 
Current peaks which could be attributed to photon assisted tunneling in 
the SET transistor were observed for bias voltages less than $4\Delta/e$.
\end{abstract}

\narrowtext

Many artificially structured microsystems such as quantum dots or
single-electron tunneling transistors have characteristic energies 
on the order of 1 meV. The frequencies needed to probe these energies 
spectroscopically are tens to hundreds of GHz. This is an 
inconvenient frequency range since there are few tunable sources 
of radiation available in this range which can be coupled to a 
sample at low temperature while isolating the sample from room 
temperature blackbody radiation.

One suitable radiation source is a Josephson junction. When 
a Josephson junction is biased above the critical current, 
voltage oscillations appear with a frequency $f = 2eV_{dc}/h$, 
where $V_{dc}$ is the dc voltage across the junction. By 
adjusting the average voltage across the junction, the 
frequency of these oscillations can be tuned. We have used 
such Josephson oscillators to measure photon assisted tunneling 
in a superconducting single electron tunneling (SET) transistor. 

In the experiments, two Al/AlO$_x$/Al Josephson junctions were 
used in parallel as in a superconducting quantum interference 
device (SQUID). Normally SQUIDs are used to measure magnetic 
flux but in this case we used the flux modulation of the critical 
current of the SQUID to tune the power of the radiation that was 
generated. The SQUID loop formed by the two junctions was 10 
$\mu$m$^2$ resulting in an inductance of about 3 pH. The 
Josephson junctions had critical currents of 1 $\mu$A and 
were shunted by a 30 $\Omega$ Pt shunt. Figure 1a shows a schematic
diagram of the circuit and Fig. 1b is an electron 
microscope image of the sample showing the coupling capacitors $C_1$ 
and $C_2$ and the positions of the 
SET transistor and the SQUID. The generator was located just 10 $\mu$m 
from the SET transistor.

It was possible to tune the frequency of the Josephson oscillations 
continuously from 10 GHz to 190 GHz.  An essential feature of this 
experiment 
is that the entire circuit is smaller than a wavelength at all 
relevant frequencies. This means that the coupling of the radiation 
from the generator to the sample circuit was not strongly
frequency dependent as is typically the case with larger 
circuits where microwave resonances affect the coupling. Another 
advantage of the small size of the circuit is that the entire 
circuit could be placed in a microwave tight enclosure with 
filtered dc feedthroughs. This prevents blackbody radiation in 
the GHz range from interfering with the experiment. The frequency 
and amplitude of the SQUID generator were adjusted with dc signals 
and the detection current flowing through the SET was also a 
dc signal.

Because the impedance of the SQUID is much smaller than the 
resistance of the SET transistor, the SQUID generator can best 
be modeled as an rf voltage source with a source impedance 
equal to the resistive shunt. The SQUID generates an rf voltage,
\cite{likharev86} 
\begin{equation}
V_{rf}(t)=\frac{R\left( I^{2}-I_{c}^{2}\right) }{I-I_{c}\sin (\omega t)}.
\label{eqn1}
\end{equation}
Here $I_c$ is the critical current of the SQUID, $I$ is the SQUID bias 
current, $R$ is the shunt resistance, and 
$\omega =\frac{2eI_{c}R}{\hbar }\sqrt{\left( I/I_{c}\right) ^{2}-1}$.

The SET transistor consisted of two low capacitance Al/AlO$_x$/Al 
tunnel junctions in series. Since these junctions were much smaller 
than the junctions in the SQUID, it was necessary to fabricate the 
SET junctions in a separate fabrication step from the SQUID junctions.
\cite{visscher95} The junctions had capacitances of 
$C_1$ = 52 aF and $C_2$ = 95 aF, resistances of $R_1$ = 400 
k$\Omega$ and $R_2$ = 90 k$\Omega$, and a gate capacitance of 
$C_g$ = 7.3 aF. A contour map of the current through the SET
transistor as a function of gate voltage and bias voltage in the 
absence of applied radiation is shown in Fig. 2. Very
little current was observed for bias voltages less than $4\Delta /e$. 
No supercurrent was observed because of the high resistance 
of the sample and the Josephson-quasiparticle cycle\cite{fulton89} 
was not observed because $e/C_{\Sigma} > 4\Delta/e$. Here
$C_{\Sigma} = C_1 + C_2 + C_g$ 
\begin{figure} [htb]
\epsfig{figure=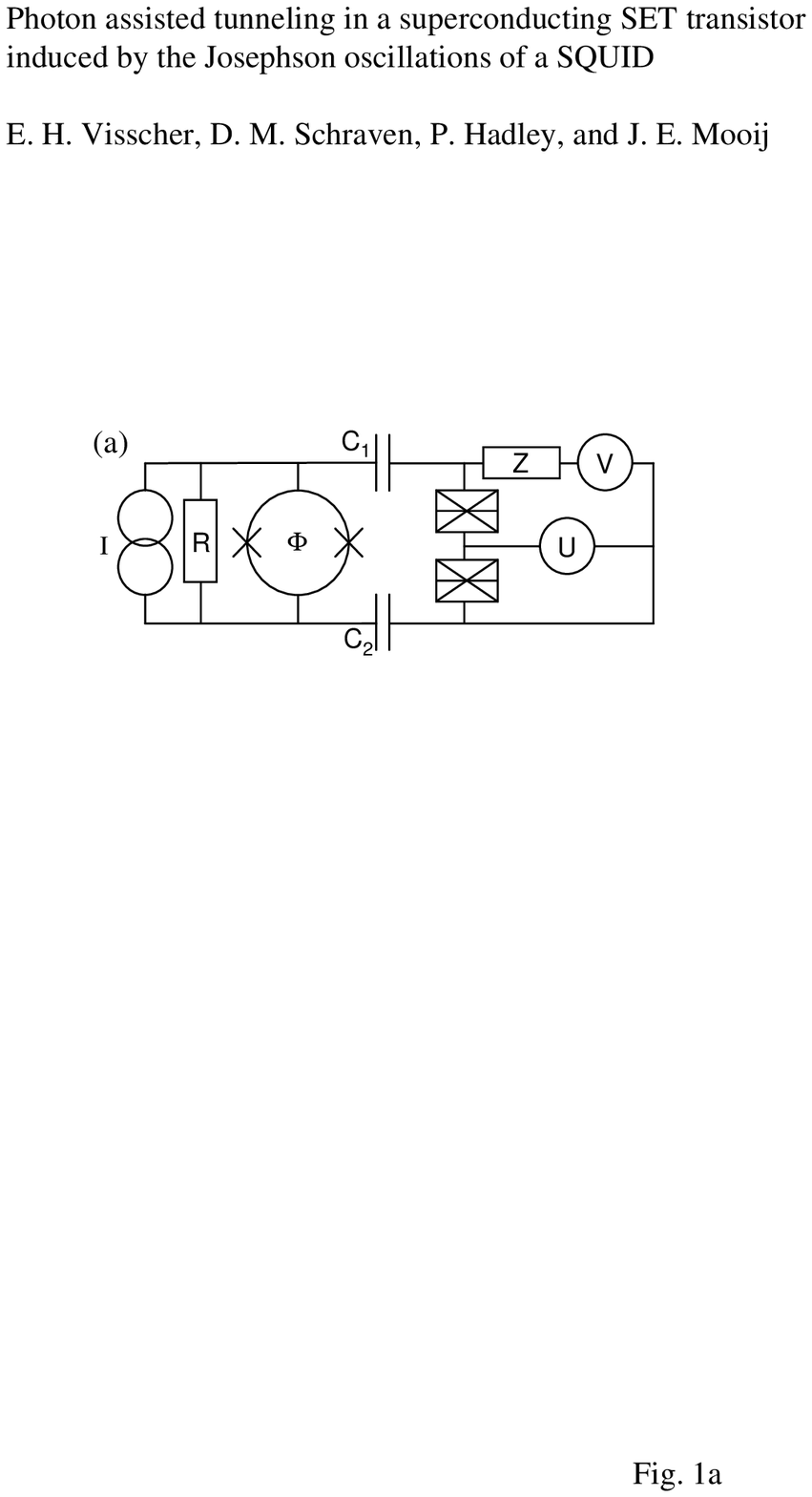, width=20.5pc, clip=true}
\epsfig{figure=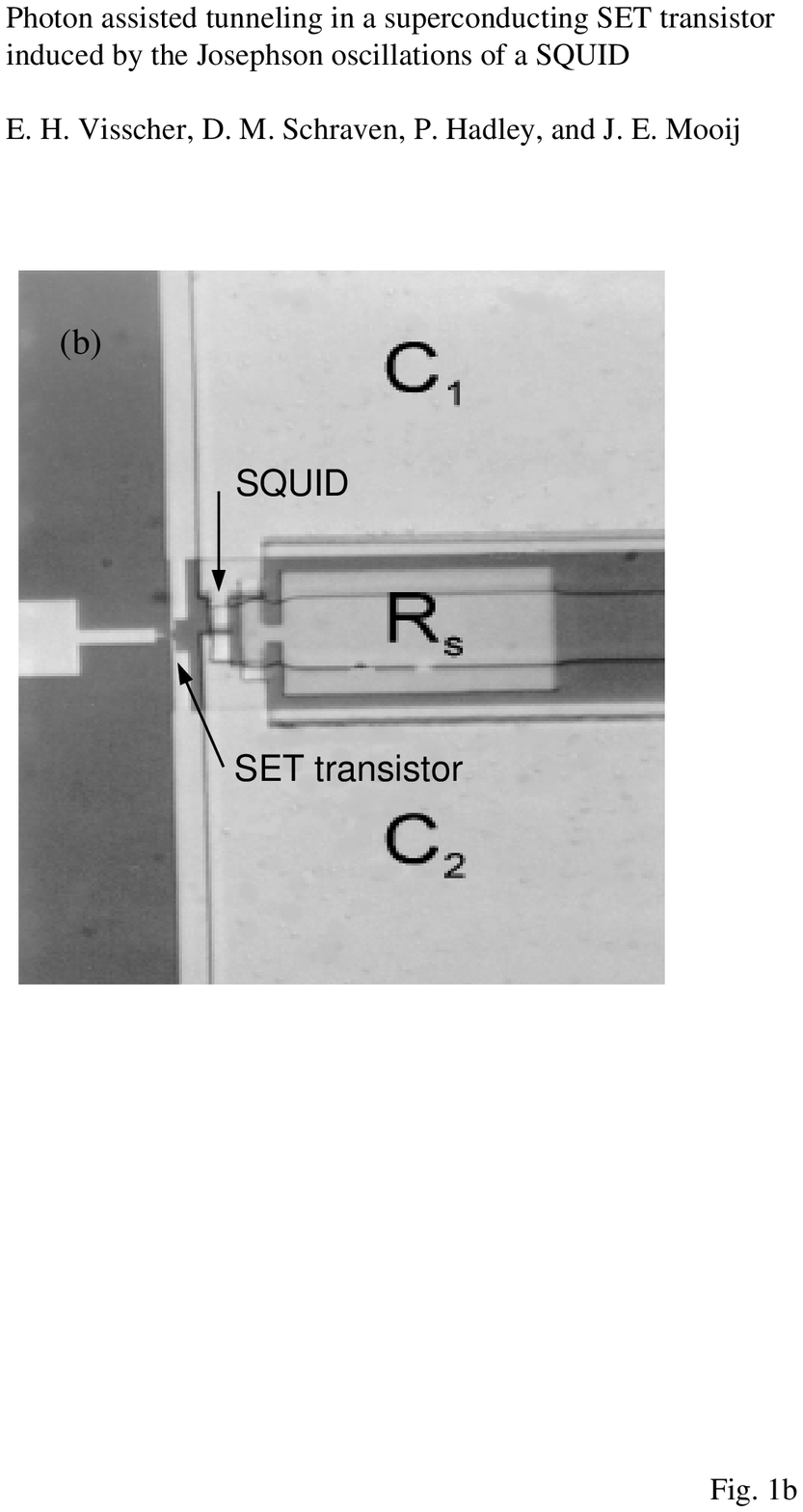, width=20.5pc, clip=true}
\caption{(a) A schematic diagram of the circuit. The SQUID is capacitively
coupled to the SET transistor. The impedance of the wires (Z) between the 
sample and the room temperature voltage source is a low impedance for low 
frequencies and high for high frequencies. (b) An electron microscope
image shows the coupling capacitors $C_1$ and $C_2$ and the positions of the 
SET transistor and the SQUID. The island 
of the SET transistor is barely visible at the point where the 
gate, source, and drain leads converge at the left of the image. 
Just to the right of the SET transistor are the two Josephson 
junctions in parallel that form the SQUID. In parallel with the 
junctions that form the SQUID is the Pt shunt resistor. The large 
appendage to the resistor is a cooling fin to keep the electron 
temperature as low as possible. The two large areas labeled 
$C_1$ and $C_2$ are the overlap capacitors that serve to couple 
the radiation generated by the SQUID to the SET transistor. The 
coupling capacitors were 200 $\mu$m $\times$ 200 $\mu$m with a 50 nm 
thick SiO dielectric. The generator was located about 10 $\mu$m 
from the SET transistor.}
\label{SEMphoto}
\end{figure}
\noindent is the total capacitance of the 
island. There is a fairly sharp onset of current that corresponds
to the threshold for sequential quasiparticle tunneling in the SET
transistor. This threshold forms the
zigzag pattern that appears between $4\Delta /e$ and $4\Delta /e
+ e/C_{\Sigma}$ in Fig. 2. 

When a bias voltage is applied across the SQUID,
\begin{figure} [htb]
\epsfig{figure=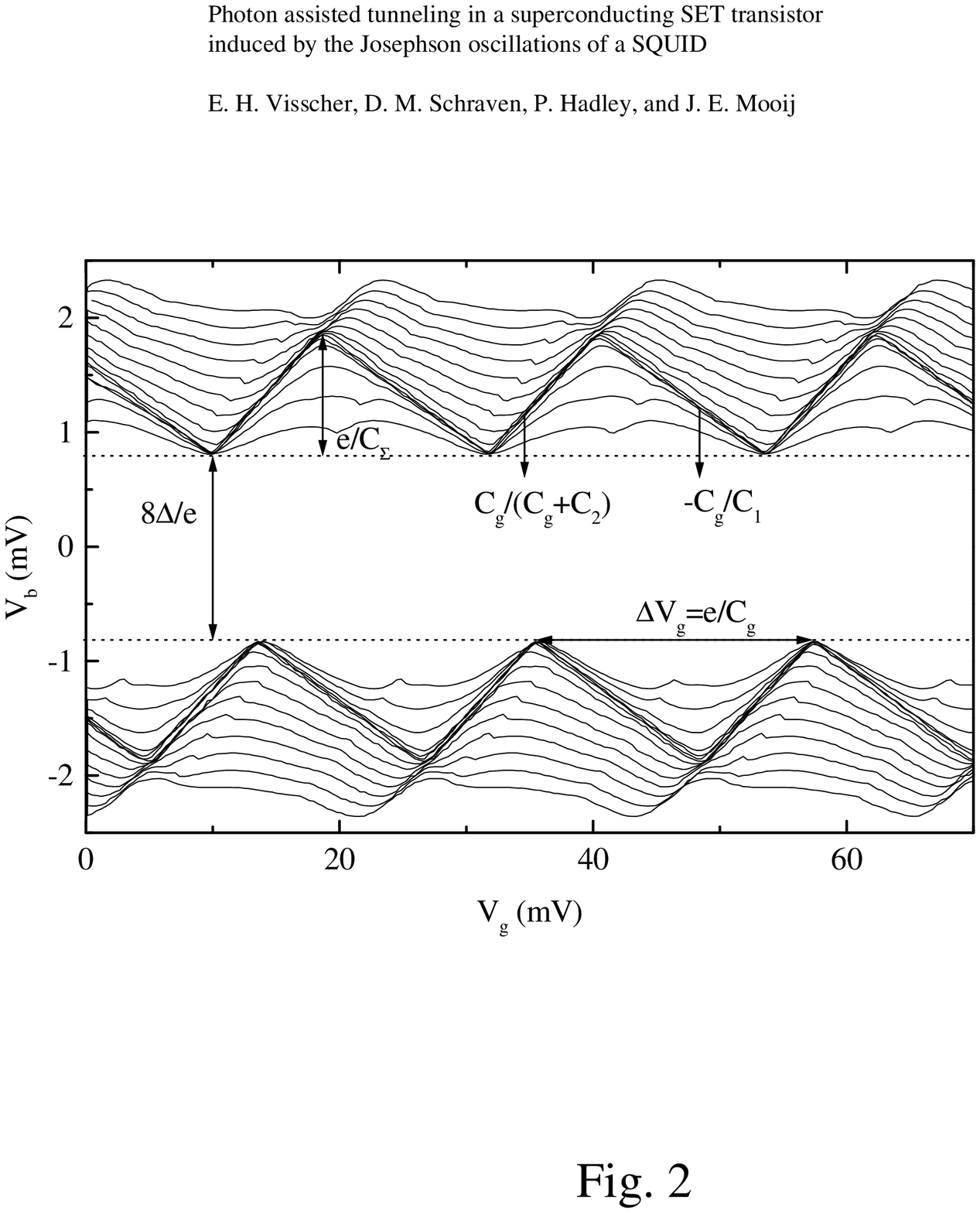, width=20.5pc, clip=true}
\caption{The bias voltage is plotted against the gate voltage for 
a variety of bias currents in the absence of rf radiation. 
From these curves the capacitances of the
junctions and the gate, and the value of the superconducting gap can be 
obtained. The currents range from -2.2 nA (lowest curve) to 2.2 nA (highest 
curve) in steps of 0.2 nA. At these current levels the current is
predominently due to the sequential tunneling of quasiparticles.}
\label{stability}
\end{figure}
\noindent Josephson 
oscillations occur which enable additional tunnel processes through 
the SET transistor via photon assisted tunneling.
\cite{nakamura96,fitzgerald} Figure 3 shows 
the current through the SET transistor when 55 GHz radiation is 
applied. The threshold for sequential quasiparticle tunneling 
corresponds to the sharp rise of current at bias voltages at 0.8 mV.
The two higher plateaus arise from photon assisted tunneling through 
one of the junctions and then direct tunneling through the other 
junction. The lower plateau is due to photon assisted tunneling through
both junctions. 

Figure 4 shows the positions of the thresholds for photon assisted
tunneling for four frequencies (0 GHz, 30 GHz, 79 GHz, and 160 GHz) in the 
$V_b - V_g$ plane. Currents above 100 pA are 
plotted as contours and the positions of smaller current peaks
and thresholds are plotted as solid points. The gray diamond shaped 
regions indicate where each quasiparticle is assisted by one photon
in tunneling through each junction. This corresponds to the lower 
diamond-shaped plateau in Fig. 3. 
A constant voltage across junction 1 
corresponds to a line with a slope of $-C_g/C_2$ in the $V_b - V_g$ 
plane and a constant voltage across junction 2 correspond to a 
line with a slope of $C_g/(C_1+C_g)$ in the $V_b - V_g$ plane. For 
this reason the thresholds all follow lines with these slopes. The 
positions of the photon assisted tunneling thresholds 
correspond to a constant voltage across one of the junctions 
that is $nhf/e$ less than the voltage at which the original threshold 
occurs. Here $n$ is an integer that indicates the order of the 
photon assisted tunneling peak. 

Figure 4a shows the current when no 
microwave radiation is applied. In addition to the threshold for
the 
\begin{figure} [htb]
\epsfig{figure=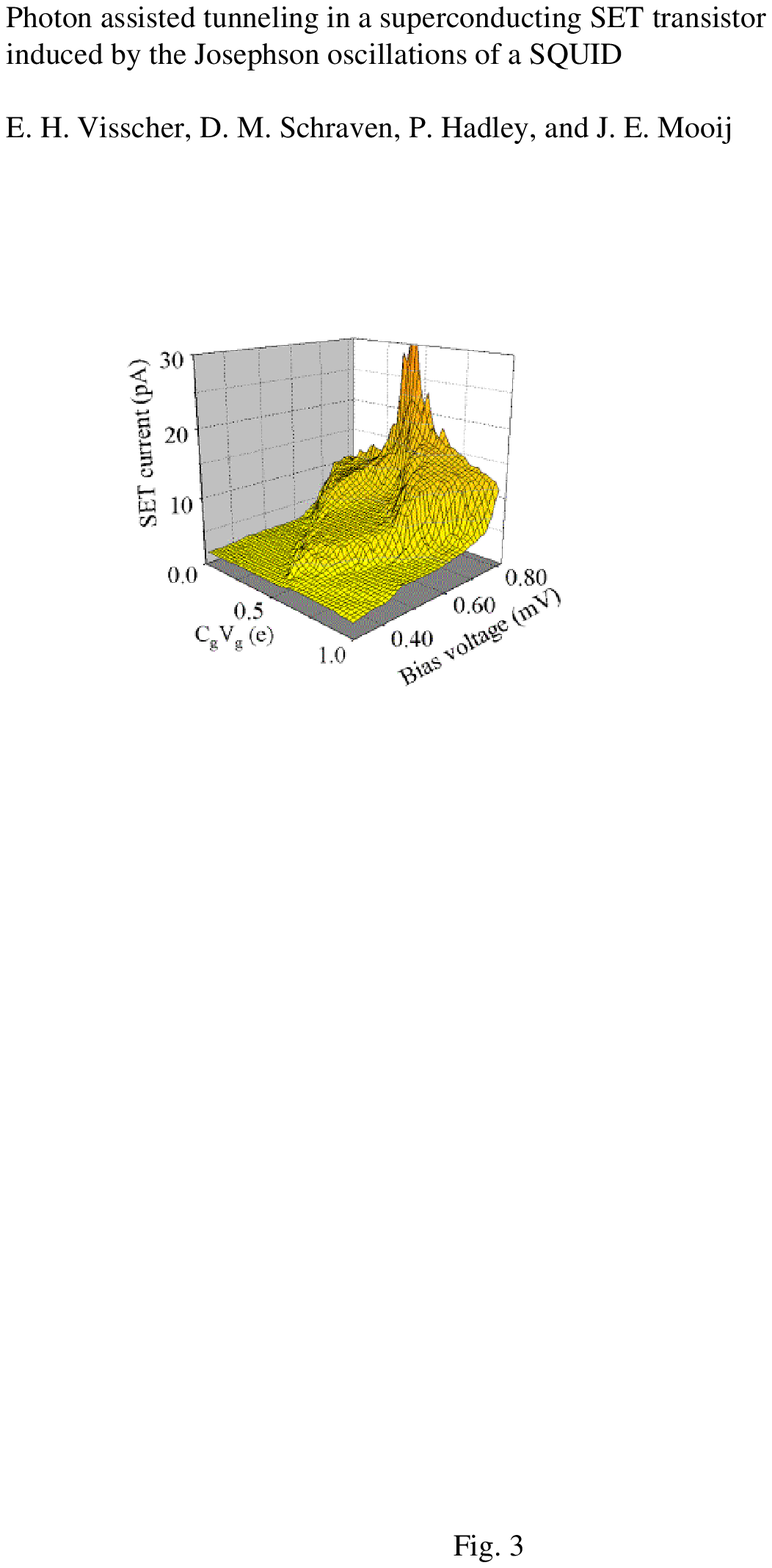, width=20.5pc, clip=true}
\caption{The current through the SET transistor as a function of
bias voltage and gate voltage when 55 GHz radiation is applied. The
sharp rise in current is the superconducting gap. The higher plateaus
correspond to photon assisted tunneling through one junction and normal
rate tunneling through the other junction. The lower plateau corresponds
to photon assisted tunneling through both junctions.}
\label{55GHz}
\end{figure}
\noindent sequential tunneling of quasiparticles, small current peaks
occur when the BCS density of states of the 
island and one of the leads aligns. These current
peaks arise from the tunneling of thermally activated quasiparticles
in a process known as 
singularity matching\cite{nakamura97} and are indicated by 
the triangles. 
A line of current peaks was also observed when the voltage across
one of the junctions equals $2\Delta /e$. If the bias across a junction 
is greater than $2\Delta /e$, then tunneling can proceed through that 
junction at the normal rate. These current peaks are indicated by 
the circles. 

Figures 4b-4d show the current through the SET transistor when high 
frequency radiation from the SQUID is applied. Again the solid
triangles represent the singularity matching peaks and the circles
represent the transition to normal 
rate tunneling in one of the two junctions. The
solid squares represent the thresholds for photon assisted tunneling. 
These squares line along the dotted lines which indicate 
the positions of thresholds for photon assisted tunneling that are
expected for the frequency of the radiation applied. The thresholds
for sequential quasiparticle tunneling and for singularity matching
do not move as a function of frequency but the thresholds for 
photon assisted tunneling scale linearly with frequency.

Photon assisted tunneling should also show a characteristic 
scaling as a function of rf power. The amplitude of the $n$th 
order peak should modulate as $J_n^2\left( \frac{eV_{rf}}{hf}\right) 
$, where $J_n(x)$ is the $n$th order Bessel function.\cite{tien63} In the
experiment, the 
zeroth order peak decreases and the higher order peaks increase as 
the power increases as would be expected from the Bessel function 
relations at low rf 
\begin{figure}  [htb]
\epsfig{figure=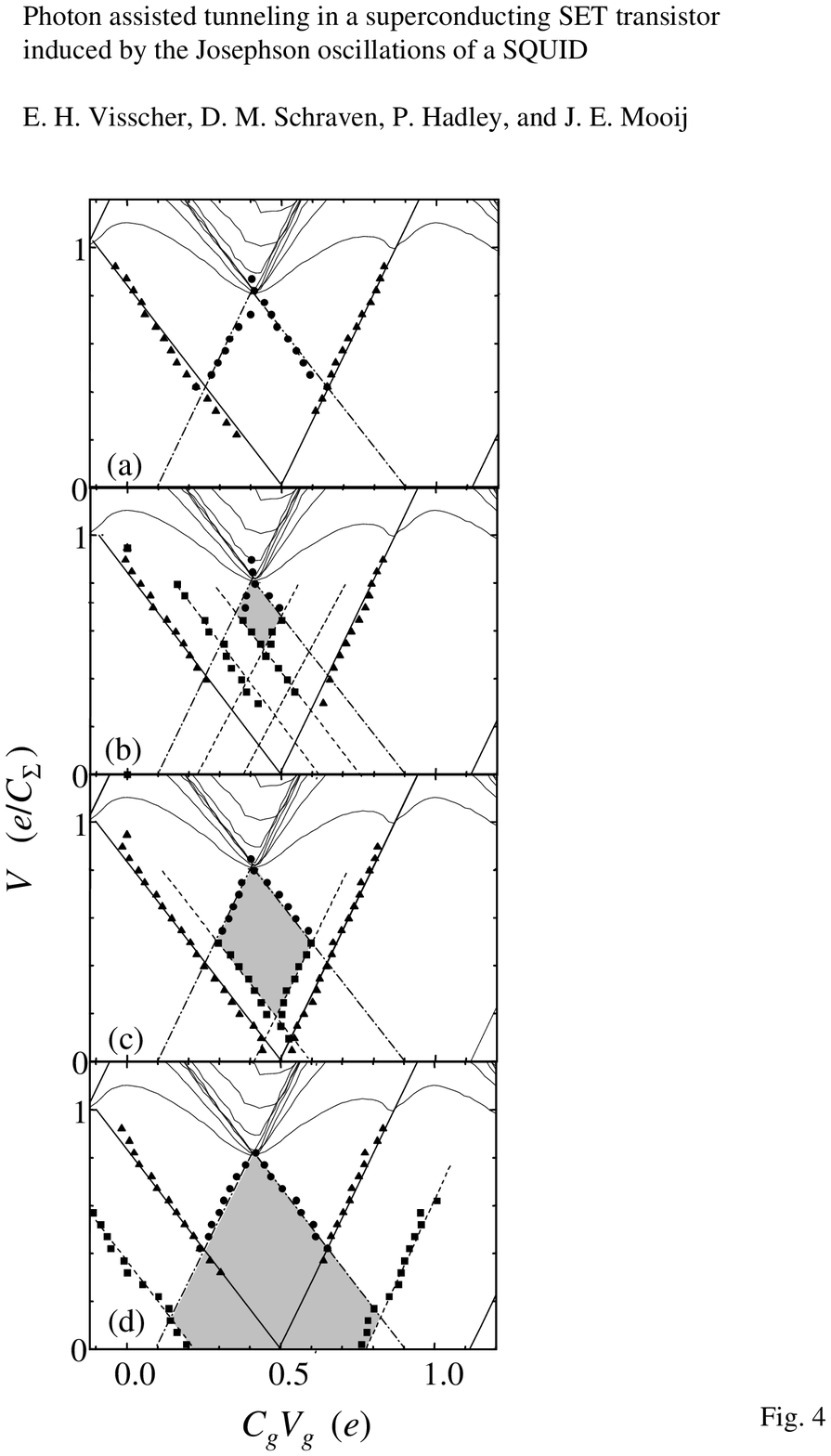, height=170mm, clip=true}
\caption{The positions of the photon assisted tunneling thresholds are
plotted for four cases of applied radiation (a) no rf, (b) 30 GHz, 
(c) 79 GHz, (d) 160 GHz. The triangles indicate the
positions of current peaks due to singularity matching. The solid lines 
indicate where singularity matching is expected to occur. The circles 
indicate current peaks that appear when the voltage across one of the
junctions equals $2\Delta /e$. The dash-dot lines indicate the positions
where this bias condition occurs. The solid squares indicate the thresholds
for photon assisted tunneling. The dotted lines indicate the positions 
where photon assisted tunneling is expected given the frequency of the 
applied radiation which is determined by measuring the voltage across the
SQUID. The gray diamond shaped regions are where each quasiparticle
is assisted by one photon in tunneling through each junction.}
\label{PAT}
\end{figure}
\noindent powers. However, the power generated by the 
SQUID was not sufficient to see the amplitude of the peaks oscillate as 
one would expect for high rf powers. From Eqn. 1, one can estimate 
that the amplitude of the rf signal should be approximately 
$V_{rf}\approx I_cR$ = 30 $\mu$V. The maximum amplitude of the rf voltage 
that a SQUID can deliver is limited by the $I_cR$ product of 
the Josephson junctions. Since the $I_cR$ product for overdamped
aluminum junctions is limited to about 0.1 mV, these Josephson generators 
will not generate enough
power to observe oscillations in the amplitude of the photon 
assisted tunneling peaks for frequencies higher than about 3 GHz. 
Higher microwave powers could be achieved by using arrays of 
junctions or other materials. 
Typical $I_cR$ products for overdamped niobium or 
high $T_c$ junctions are about 1 mV. This leads to oscillations in 
the amplitude of the photon assisted tunneling peaks for 
frequencies up to about 30 GHz for single junction oscillators. 

In conclusion, high frequency oscillations generated by a SQUID 
were used to irradiate a superconducting SET transistor. 
This radiation induced photon assisted tunneling in the SET transistor 
over a span of frequencies from 10 GHz to 190 GHz. 
A linear scaling
with the frequency was observed. An advantage of using SQUIDs as the
rf source is that microwave resonances in the measurement circuit
were not observed so that it was possible to tune the frequency 
continuously. 
Placing the rf oscillator on the same chip as the sample to be measured
also makes it possible to enclose the entire measurement circuit in 
a microwave tight box which minimizes external interference such as
the blackbody radiation of room temperature equipment. 

\acknowledgments

Support from Esprit project 22953, CHARGE, is 
gratefully acknowledged.

\widetext
\end{document}